\documentclass{aa}

\usepackage{amsmath}
\usepackage{bm}
\usepackage{graphicx}
\usepackage{natbib}
\usepackage{placeins}
\usepackage{subcaption}
\usepackage[flushleft]{threeparttable}
\usepackage{txfonts}
\usepackage{upgreek}
\bibpunct{(}{)}{;}{a}{}{,}

\begin{document}

   \title{Nonthermal emission from high-mass microquasar jets affected by orbital motion}
   
   \titlerunning{Nonthermal emission from HMMQ jets affected by orbital motion}

   \author{E. Molina \and V. Bosch-Ramon}

   \institute{Departament de Física Quàntica i Astrofísica, Institut de Ciències del Cosmos (ICCUB),
              Universitat de Barcelona (IEEC-UB), Martí i Franquès 1, 08028 Barcelona, Spain \\
              \email{emolina@fqa.ub.edu}}

   \date{Received -/Accepted -}

  \abstract 
   {The stellar wind in high-mass microquasars should interact with the jet. This interaction, coupled with orbital motion, is expected to make the jet follow a helical, nonballistic trajectory. The jet energy dissipated by this interaction, through shocks for example, could lead to nonthermal activity on scales significantly larger than the system size.}
   {We calculate the broadband emission from a jet affected by the impact of the stellar wind and orbital motion in a high-mass microquasar.}
   {We employ a prescription for the helical trajectory of a jet in a system with a circular orbit. Subsequently, assuming electron acceleration at the onset of the helical jet region, we compute the spatial and energy distribution of these electrons, and their synchrotron and inverse Compton emission including gamma-ray absorption effects.}
   {For typical source parameters, significant radio, X- and gamma-ray luminosities are predicted. The scales on which the emission is produced may reduce, but not erase, orbital variability of the inverse Compton emission. The wind and orbital effects on the radio emission morphology could be studied using very long baseline interferometric techniques.}
   {We predict significant broadband emission, modulated by orbital motion, from a helical jet in a high-mass microquasar. This emission may be hard to disentangle from radiation of the binary itself, although the light curve features, extended radio emission, and a moderate opacity to very high-energy gamma rays, could help to identify the contribution from an extended (helical) jet region.}

   \maketitle

\section{Introduction}\label{intro}

High-mass microquasars (HMMQ) are X-ray binaries that host a massive star and a compact object (CO) from which jets are produced. The stellar wind can strongly influence the jet propagation, both because the jet has to propagate surrounded by wind material, and because the wind lateral impact may significantly bend the jet away from the star. Several authors have used numerical and analytical methods to study, on the scales of the binary, the interaction of HMMQ jets with stellar winds and their radiative consequences \citep[e.g.,][]{romero03,romero05,perucho08,owocki09,araudo09,perucho10,perucho12,yoon15,bosch16,yoon16}. 

On scales larger than the binary system, orbital motion should also affect the dynamics of the jet, making it follow a helical trajectory \citep[e.g.,][]{bosch13,bosch16}. There are a few well-established microquasars in which evidence of a helical jet has been observed; for example, SS 433, 1E~1740.7$-$2942 and Cygnus~X-3 (\citealt{abell79,mioduszewski01,stirling02,miller04,luque15}; see also \citealt{sell10} for the possible case of Circinus~X-1). In the case of SS 433, the jet helical geometry likely originates in the accretion disk \citep[e.g.,][]{begelman06}; in the case of 1E~1740.7$-$2942, the system likely hosts a low-mass star and the role of its wind could be minor; in the case of Cygnus~X-3, the apparent jet helical shape has an unclear origin. \cite{bosch16} proposed that Cygnus~X-1 and Cygnus~X-3 could be affected by orbital motion and the stellar wind, although the uncertainties on the jet and wind properties make quantitative predictions difficult. Unlike the case of Cygnus~X-3, no clear evidence of a helical or bent jet has yet been found for Cygnus~X-1 \citep{stirling01}.

In this work we study the implications of the impact of the stellar wind and orbital motion for the nonthermal emission of a HMMQ, from radio to gamma rays. Using an analytical prescription for the jet dynamics based on the results of \cite{bosch16}, the nonthermal radiation from a helical jet (spectra, light curves, and morphology) is computed numerically, considering a leptonic model with synchrotron and inverse Compton (IC) emission. 

The paper is organized as follows: In Sect.~\ref{windjet}, the interaction between the stellar wind and the jet is described. The technical aspects of the model are explained in Sect.~\ref{model}. The results of the calculations are presented in Sect.~\ref{results} and a discussion is provided in Sect.~\ref{discussion}. Unless stated otherwise, the convention $Q_{\rm x} = Q/10^{\rm x}$ is used throughout the paper, with $Q$ in cgs units.

\section{Jet-wind interaction}\label{windjet}

The physical scenario considered in this work consists of a massive star with a strong stellar wind and an accreting CO from which two jets are launched in opposite directions perpendicular to the orbital plane. The binary has a relatively compact orbit, taken circular for simplicity, with a period of $P = 5$~days and a separation of $d_{\rm O} = 3\times10^{12}$~cm ($M_1+M_2 \approx 43~M_\sun$). The luminosity of the star is taken to be $L_\star = 10^{39}$~erg~s$^{-1}$, and its temperature $T_\star = 4\times10^4$~K ($R_\star \approx 10 \ R_\sun$). The stellar wind is assumed to be spherically symmetric, with a mass-loss rate and a velocity of $\dot{M}_{\rm w} = 10^{-6}$~M$_\sun$~yr$^{-1}$ and $v_{\rm w} = 2\times10^8$~cm~s$^{-1}$, respectively \citep[typical for O-type stars; e.g.,][]{muijres12}. Such a system would have similar parameters to those of Cygnus~X-1, although it would be significantly wider and with a weaker wind than Cygnus~X-3 (but the adopted wind and jet momentum rate relation will be similar; see below).

The orbit is assumed to lie on the $xy$-plane, with the star at the center of coordinates. The CO would be on the $x$-axis, orbiting the star counter clockwise, and the jet (counter jet) would  initially point along $\hat z$ ($-\hat z$). The axes are defined in a frame corotating with the CO around the companion star. The jets are assumed to have a conical shape, with an initial half-opening angle $\theta_{\rm j0} = 0.1$~rad~$\approx 5.7^\circ$. In this work, we do not need to adopt a specific value for $\theta_{\rm j0}$, although to make our predictions qualitatively valid, it should be smaller than the jet deflection angle imposed by the wind impact (see below). We take an initial jet Lorentz factor of $\gamma_{\rm j} = 2$, although the final results are not sensitive to this parameter as long as the jet is strongly supersonic and mildly relativistic. The jet power is taken as $L_{\rm j} = 3\times10^{36}$~erg~s$^{-1}$.

\begin{figure}
        \centering
        \includegraphics[width=\linewidth]{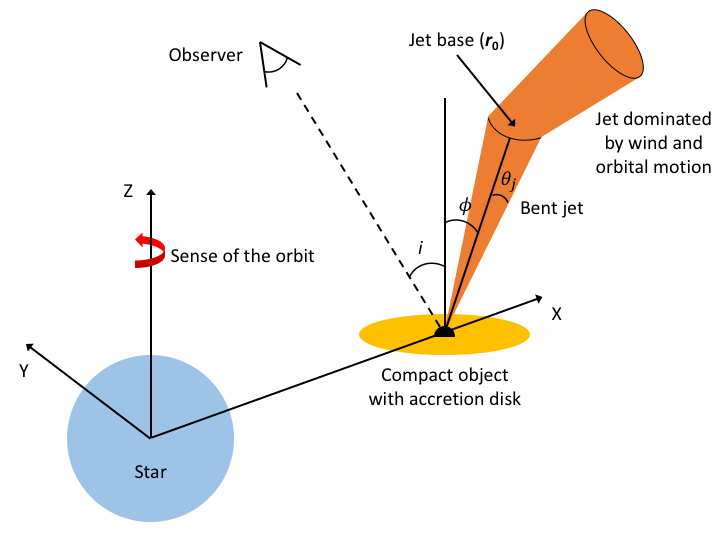}
        \caption{Schematic representation of the onset of the helical jet. The position, with respect to the star, of the change in the jet initial (bent) direction is $\boldsymbol{r_0}$, where the jet bending starts to be dominated by the added effects of the stellar wind and orbital motion. The axes are defined in a frame corotating with the CO.}
        \label{fig:windjet}
\end{figure}

Figure~\ref{fig:windjet} sketches the scenario under study. The jet goes through three stages, the first of which is not present in the figure: Initially, the jet moves perpendicularly to the orbital plane. Then, the wind impact produces an asymmetric recollimation shock on the jet, which gets inclined away from the star (stage 2). Further from the binary, the combined effect of the stellar wind and orbital motion leads to a (orbit-related) force that bends the already inclined jet, now against orbital motion (clockwise in Fig.~\ref{fig:windjet}). In this way, the jet becomes helical, with asymmetric shocks being expected because of the orbit-related force (stage 3). Here it is assumed that electrons are efficiently accelerated in those shocks, ascribed here to one specific point: the onset of the helical jet, which neglects the fact that acceleration may occur along the helical jet region. Relativistic protons may also be generated in those shocks, but their radiation efficiency would be much lower in general, and therefore they are not considered in our calculations \citep[see e.g.,][for a discussion of the different electron and proton cooling timescales in HMMQ]{bosch09a}.

For the radiation calculations, a fraction $\eta_{\rm NT} = 10^{-2}$ of $L_{\rm j}$ is injected in the form of nonthermal electrons right where the jet starts its helical path. This fraction is only constrained by $\eta_{\rm NT} < 1$, although the predicted luminosities are proportional to its value and easy to translate if a different fraction is adopted. Nevertheless, adopting $\eta_{\rm NT} = 10^{-2}$ does not require particularly efficient acceleration processes. The inclination of the system with respect to the line of sight of the observer, assumed to be at a reference distance of $d = 3$~kpc, is characterized by the angle $i$. The values for the different fixed parameters of our model are listed in Table~\ref{tab:parameters}.

\begin{table}
 \begin{center}
	\caption{List of fixed parameters used in this work.}
	\begin{tabular}{l c c}
    \hline \hline
	\multicolumn{1}{c}{Parameter}		&					&	Value							\\
    \hline
		Star temperature				& $T_\star$			&	$4\times10^4$ K 				\\
        Star luminosity 				& $L_\star$ 		& 	$10^{39}$ 	erg~s$^{-1}$		\\
        Wind speed 						& $v_{\rm w}$ 		&	$2\times10^8$~cm~s$^{-1}$		\\
        Wind mass-loss rate 			& $\dot{M}_{\rm w}$ &	$10^{-6}$ M$_\sun$~yr$^{-1}$	\\
    	Jet luminosity 					& $L_{\rm j}$ 		&	$3\times10^{36}$ erg~s$^{-1}$	\\
        Initial jet Lorentz factor 		& $\gamma_{\rm j}$ 	&	2								\\
        Half-opening angle 				& $\theta_{\rm j}$ 	&	0.1 rad							\\
        Nonthermal energy fraction 		& $\eta_{\rm NT}$ 	&	$10^{-2}$						\\
        Effective flow velocity			& $v_{\rm j}$		&	$5\times10^9$ cm~s$^{-1}$		\\
        Radial flow velocity			& $v_{\rm r}$		&	$2\times10^9$ cm~s$^{-1}$		\\
    	Orbital separation 				& $d_{\rm O}$ 		&	$3\times10^{12}$ cm 			\\ 	
    	Period 							& $P$ 				&	5 days							\\
        Distance to the observer 		& $d$ 				&	3 kpc 							\\ 	
    	Radio interferometer resolution	& FWHM				&	1 mas							\\
        \hline
    \end{tabular}
    \label{tab:parameters}
 \end{center}
\end{table}

\subsection{Wind impact: jet bending}\label{wi}

Interaction between the stellar wind and the jet can lead to bending of the latter in the $x$-axis direction (stage 2 above). This bending can be characterized by an angle $\phi$ with respect to the jet initial direction (i.e., the $z$-axis; see Fig.~\ref{fig:windjet}). Following \cite{bosch16} we introduce a nondimensional parameter, $\chi_{\rm j}$, which corresponds to the ratio between the wind momentum rate intercepted by the jet, and the jet momentum rate:
        \begin{equation}\label{chi}
        \chi_j \approx \frac{\theta_{\rm j0} \dot{P}_{\rm w}}{4\pi \dot{P}_{\rm j}} =
               \frac{\theta_{\rm j0} \dot{M}_{\rm w} v_{\rm w} (\gamma_{\rm j}-1) c}
               {4\pi L_{\rm j} \gamma_{\rm j} \beta_{\rm j}}
               \approx 0.9 \frac{\theta_{\rm j0,-1} \dot{M}_{\rm w,-6} v_{\rm w,8.3}}
               {L_{\rm j,36.5} \beta_{\rm j}} \frac{(\gamma_{\rm j}-1)}{\gamma_{\rm j}} \ ,
        \end{equation}
where $\dot{M}_{\rm w}$ is in units of M$_\sun$~yr$^{-1}$ in the rightmost expression, $\dot{P}_{\rm w} = \dot{M}_{\rm w} v_{\rm w}$ is the stellar wind momentum rate, and $\dot{P}_{\rm j}$ is the jet momentum rate, which can be expressed in terms of $L_{\rm j}$, $\gamma_{\rm j}$, and the initial jet velocity in $c$ units $\beta_{\rm j0}$:
        \begin{equation}\label{Pj}
        \dot{P}_j = \frac{L_{\rm j} \gamma_{\rm j} \beta_{\rm j0}}{c (\gamma_{\rm j}-1)} \ .
        \end{equation}

An approximate computation of the bending angle $\phi$, with a discrepancy of less than 10\% with respect to the numerical case, yields the following expression (for more details, see the Appendix in \citealt{bosch16}):
        \begin{equation}\label{phi}
        \phi = \frac{\pi^2 \chi_{\rm j}}{2\pi\chi_{\rm j} + 4\chi_{\rm j}^{1/2} + \pi^2} \ .
        \end{equation}

For bending angles $\phi < \theta_{\rm j0}$, the lateral impact of the stellar wind may produce a weak recollimation shock or a sound wave in the jet, and the large-scale jet evolution should not be strongly affected \citep[although the role of instability growth may still be important; see, e.g.,][under uniform and clumpy stellar winds]{perucho08,perucho10,perucho12}. However, when $\phi$ is significantly larger than $\theta_{\rm j0}$, a strong asymmetric recollimation shock is expected, with the jet substantially deviating from its initial direction. In such a case, the jet is likely to turn into a helical-shaped structure at larger scales (stage 3, above). Consequently, bending due to stellar wind impact becomes important for the jet evolution when $\phi > \theta_{\rm j0}$.

\subsection{Orbital effects: helical jet}

The orbital motion of the system combined with a jet inclined by the stellar wind (stage 2) leads, through the orbit-related force, to a jet helical pattern. This effect becomes clearer for significant bending angles. Otherwise, for $\phi < \theta_{\rm j0}$, the conical jet expansion will dominate the geometry of the resulting structure. We note that a helical pattern would be expected from an inclined jet even under ballistic jet propagation, but the presence of the stellar wind makes jet propagation nonballistic. Instability growth should also play a role, but here we consider its effects only phenomenologically through a slower jet flow, caused by the wind mass entrainment and hence deceleration associated to instabilities.

One can estimate the distance $x_{\rm turn}$ that a nonballistic jet travels in the $x$-axis direction before acquiring a negative velocity $y$-component, i.e., bending against the orbital motion. With the additional assumptions that the half-opening angle of the jet remains constant after being bent (see below), and that $x_{\rm turn}$ is much larger than the orbital separation, we get \citep[see][for the derivation of this expression]{bosch16}:
        \begin{equation}\label{xturn}
        x_{\rm turn} = \frac{K}{\sqrt{2\chi_{\rm j}}}  \frac{v_{\rm w}}{\omega} \approx
                       1.3\times10^{13} \frac{K}{\sqrt{\chi_{\rm j}}}
                       \frac{v_{{\rm w},8.3}}{\omega_{-5}} \ \ {\rm cm} \ ,
        \end{equation}
where $\omega$ is the orbital angular velocity, with the normalization corresponding to a period of several days (e.g., $\omega = 2 \pi / 5 \ {\rm days}\approx 1.4\times 10^{-5}$~rad~s$^{-1}$), and $K$ is a constant of $O(1)$, introduced to account for the specific details of the wind-jet interaction. Here we fix $K=1$.

Knowing $x_{\rm turn}$ and $\phi$, we can compute the height $z$ at which the jet starts forming the helical structure: $z_{\rm turn} = x_{\rm turn} / \tan\phi$. For the parameters given above, and following Eqs.~\eqref{phi} and \eqref{xturn}, the location vector of the beginning of the helical structure is $\boldsymbol{r_0} = (1.58, 0, 3.56)\times10^{13}$ cm, with $\phi = 19.8^\circ$. We note that $x_{\rm turn} = 1.28\times10^{13}$~cm$ \ \gg d_{\rm O}$, as required. The distance from $\boldsymbol{r_0}$ to the star is $D_\star = 3.88\times10^{13}$~cm and to the CO is $l_0 = 3.77\times10^{13}$~cm. The jet curvature due to orbital motion for distances smaller than $l_0$ is neglected, that is, $y_{\rm turn}=0$.

The point $\boldsymbol{r_0}$ (the -helical- jet base hereafter) is the position where nonthermal electrons are injected. The evolution of these electrons is followed from this point onwards along the helical jet assuming that no more significant electron acceleration happens further downstream in the jet\footnote{We note that as the jet propagation is nonballistic, (weak) acceleration regions may also exist all along the jet due to jet kinetic energy dissipation.}. Electron acceleration may also happen closer to the binary system (see Sect.~\ref{discussion}), but here we focus specifically on the nonthermal emission produced in the helical jet region. 

From $\boldsymbol{r_0}$ onwards along the jet, the stellar wind should have a very important dynamical role. As the jet becomes helical due to the presence of wind plus orbital motion, the former is also pushing the latter in the radial direction. Therefore, the radial velocity ($v_{\rm r}$) of the jet flow may be as slow as the stellar wind itself, or higher if the shocked stellar wind is effectively accelerated radially by the jet. In addition, effects of wind entrainment are likely to affect the jet flow itself, slowing the latter down to an effective velocity $v_{\rm j}<\beta_{\rm j0} c$. We adopt here $v_{\rm r}$ and $v_{\rm j}$ as phenomenological parameters that fulfill $v_{\rm w}\lesssim v_{\rm r}\lesssim v_{\rm j}$. They provide a simplified prescription for the helical jet geometry, defining the pitch of the helical jet trajectory. It is also assumed that this trajectory is confined to the conical surface, characterized, in the nonrotating frame, by the direction determined by the vector $\boldsymbol{r_0}$ along the orbit in that frame (i.e., a ring traced by the points $\boldsymbol{r_0}$, plus the radial direction from the star), which should be a relatively good approximation as long as $l_0 \gg d_{\rm O}$. 

The half-opening angle of the helical jet flow ($\theta_{\rm j}$) need not be constant, as the nonballistic nature of the jet trajectory, with weak-shock heating, rarefaction waves, wind (re)confinement, and mass-loading, may lead to (re)collimation or (re)widening of the helical jet, depending on the details of the jet and wind interaction. For simplicity, at this stage we assume $\theta_{\rm j}=\theta_{\rm j0}$. In addition, we also take $v_{\rm j}=5\times 10^9$~cm~s$^{-1}$ and $v_{\rm r}=2\times 10^9$~cm~s$^{-1}$; this is somewhat arbitrary, but except for the radio morphology (see Sect.~\ref{discussion}) their actual values do not qualitatively change the results. Hydrodynamical simulations are required to properly characterize $v_{\rm r}$, $v_{\rm j}$, $\theta_{\rm j}$, the role of instability growth, mass-loading, and so on. Nevertheless, we consider our dynamical model realistic on scales of a few $D_\star$, as large-scale perturbations should grow up  to a size $\gtrsim D_\star$ to disrupt the helical structure.

\section{Model description}\label{model}

We use a semi-analytical code to compute the energy and spatial distribution of the accelerated electrons. As the orbital period is significantly longer than the crossing time of the helical jet length ($l_{\rm max}/v_{\rm j}$), the nonthermal electrons are assumed to be in the steady state at each orbital phase.

\subsection{Energy losses}

Analytical expressions for the energy losses are used in this work. For synchrotron cooling, assuming, for simplicity, an isotropic magnetic field $B$ in the flow frame, one obtains, in cgs units \citep{longair81}:
        \begin{equation}\label{synloss}
        \dot{E}_{\rm sync} \approx -\frac{4}{3} c \sigma_{\rm T} \omega_{\rm mag} \gamma^2 \approx
    1.6\times10^{-5} B_{-2}^2 E^2 \ {\rm erg~s}^{-1},
        \end{equation}
where $\sigma_{\rm T}$ is the Thomson cross-section, $\omega_{\rm mag} = B^2 / 8\pi$, and $\gamma$ is the Lorentz factor of an electron with energy $E$. The magnetic field at the jet base is parametrized through the ratio of magnetic pressure to stellar photon energy density, $\eta_B$, as
        \begin{equation}\label{nB}
        \frac{B^2}{8\pi} = \eta_B \frac{L_\star}{4 \pi c D_\star^2} \ .
        \end{equation}
This way of normalizing $B$ allows an easy comparison between the expected luminosity of synchrotron and IC emission; for IC in the Thomson regime, the two processes contribute similarly for electrons of the same energy if $\eta_{\rm B}\sim 1$. However, for electrons of energy $\gtrsim m_{\rm e} c^2/3k_{\rm B}T_\star$ (with $k_{\rm B}$ being the Boltzmann constant), which scatter photons in the IC Klein-Nishina regime, $\eta_{\rm B}\gtrsim 0.1$ already leads to global strong synchrotron dominance, which is corroborated by our calculations. One can relate $\eta_{\rm B}$ with the ratio of magnetic pressure to jet energy density ($\bar\eta_{\rm B}$) for the same $B$ via: $\bar\eta_{\rm B}=\theta_{\rm j}^2(\eta_{\rm B}/4)(L_\star/L_{\rm j})(v_{\rm j}/c)$ ($\approx 0.14\eta_{\rm B}$ for the values adopted here).

Losses due to IC are computed as in \cite{khangulyan14} for an isotropic electron population embedded in the stellar photon field taken as a black body of temperature $T_\star$. This expression is valid in both the Thomson and the Klein-Nishina regimes and has an accuracy of $\sim \! 5\%$:
        \begin{equation}\label{ICloss}
    	\begin{split}
        \dot{E}_{\rm IC} & \approx
    	-\frac{3 \sigma_{\rm T} \pi m_{\rm e}^2 c^2 R_\star^2 k_{\rm B}^2 T_\star^2}{2 R^2 h^3} G(t) \\
    	& \approx -3.8\times10^{-3} \frac{R_{\star,12}^2 T_{\star,4.6}^2}{R_{14}^2}
    	\frac{E \ln (1+5.3E)}{1+183E} \ \ {\rm erg~s}^{-1} \ ,
    	\end{split}
        \end{equation}
with $R_\star$ being the radius of the star, $R$ the distance from the star to the emitter, $t = 4 E k_{\rm B} T_\star / m_{\rm e}^2 c^4$, and
        \begin{equation}\label{G}
    	G(t) = \frac{4.62t \ln(1+0.156t)}{1+5.62t} \ .
        \end{equation}  
        
For a conical jet, adiabatic losses are
        \begin{equation}\label{adloss}
        \dot{E}_{\rm ad} = -\frac{2}{3} \frac{v_\perp}{r_j} E \approx
                           -6.7\times10^{-6} \frac{v_{\perp,8}}{r_{j,13}} E \ \ {\rm erg~s}^{-1} \ ,
        \end{equation}
where $v_\perp \approx v_{\rm j}\theta_{\rm j}$ is the jet expansion velocity and $r_{\rm j}$ is the helical jet radius.

The total energy-loss rate is
        \begin{equation}\label{loss}
        \dot{E} = \dot{E}_{\rm sync} + \dot{E}_{\rm IC} + \dot{E}_{\rm ad} .
        \end{equation}

\subsection{Electron distribution}

The nonthermal electrons are injected in the jet base with an energy distribution that follows a power law of index -2, typical for efficient nonthermal sources \citep[and for acceleration in nonrelativistic strong shocks;][]{drury83}, plus a cutoff at high energies set by energy and escape losses:
        \begin{equation}\label{Q}
        Q(E) \propto E^{-2} \exp{\left( -\frac{E}{E_{\rm cut}}\right)} \ .
        \end{equation}
We note that a much steeper $Q(E)$ would mean inefficient electron acceleration, whereas a much harder one would imply exceptionally efficient acceleration in which most of energy is in the highest-energy electrons. 

The normalization of the electron injection is taken as a fraction of the jet power available for the acceleration of nonthermal electrons, $L_{\rm NT} = \eta_{\rm NT} L_{\rm j}$:
        \begin{equation}\label{normQ}
        \int_{E_{\rm min}}^{E_{\rm max}} \! E \ Q(E) \ {\rm d} E = L_{\rm NT} \ .
        \end{equation}
The value of $E_{\rm min}$ is fixed here to 1~MeV for simplicity, although a very high $E_{\rm min}$, greater than $0.1-1$~GeV for example, would affect the more compact radio emission. The value of $E_{\rm max}$ is obtained from electron acceleration constraints.

Electrons are accelerated gaining energy at a rate $\dot{E}_{\rm acc} = \eta_{\rm acc} e B c$, where $\eta_{\rm acc}$ is the acceleration efficiency, and $e$ the electron charge. We take $\eta_{\rm acc} = 0.01$, which would correspond, for example, to acceleration by a high-velocity (nonrelativistic) shock under Bohm diffusion \citep{drury83}. Low acceleration efficiencies would reduce the maximum energy of synchrotron photons, for which the photon energy is $\varepsilon\propto E^2$, much more than that of IC photons, produced in the Klein-Nishina regime at the highest spectral end, for which $\varepsilon \sim E$ (for IC in Thomson: $\varepsilon\propto E^2$).

Equating $\dot{E}_{\rm acc}=|\dot{E}|$ yields the maximum energy that electrons can reach in the accelerator (at $\boldsymbol{r_0}$): $E_{\rm acc}$. However, electrons can diffuse away from the accelerator before they reach $E_{\rm acc}$. Therefore, the diffusion timescale, $t_{\rm d} = r_{\rm j0}^2 / 2D_{\rm B}$, is also to be compared with the acceleration timescale, $t_{\rm acc} = E_{\rm d} / \dot{E}_{\rm acc}$, where $r_{\rm j0}$ is the characteristic jet base radius, and $D_{\rm B} = E_{\rm d} c / 3qB$ is the diffusion coefficient assuming it proceeds in the Bohm regime. This gives a maximum energy value $E_{\rm d} = q B r_{\rm j} \sqrt{3 \eta_{\rm acc}/2}$; for a larger diffusion coefficient $D$ (still $\propto E$), $E_{\rm d}\propto \sqrt{D_{\rm B}/D}$. The $E_{\rm cut}$ value is obtained as the smallest among $E_{\rm acc}$ and $E_{\rm d}$, whereas $E_{\rm max}$ can be taken as several times $E_{\rm cut}$.

To compute the electron energy distribution along the jet, the latter is divided in cylindrical segments with increasing radius. The segment lengths are determined (with some exception; see below) by $v_{\rm j}$ multiplied by one fifth of the local shortest cooling time, which is derived from $t_{\rm loss} = E_{\rm max}/|\dot{E}(E_{\rm max})|$. In this way, the energy and spatial evolution of the fastest evolving electrons is reasonably well sampled. The magnetic field is assumed to be mostly perpendicular to the flow motion, typical for jets far from their origin, and therefore $B\propto r_{\rm j0}/r_{\rm j}$ under frozen conditions and constant $v_{\rm j}$. Each individual jet division is treated as a homogeneous emitter, which in the worse case is correct within a $\sim 3$\% error (spatial scales grow segment by segment by a factor $1+l_{\rm max}/N/l_0\sim 1.03$, where $N$ is the number of segments).

For $\eta_{\rm B}\rightarrow 1$, synchrotron cooling times become very short ($t_{\rm loss} \ll t_{\rm adv}$, with $t_{\rm adv}$ being the advection timescale), and the large $N$ makes calculations very long. In those extreme cases, we have adopted an approximation which consists in limiting $N$ and then introducing a correction factor, $\times \ t_{\rm loss}/t_{\rm adv}$, to the injected population at the first segment. This simplified approach can overestimate the high energy part of the electron distribution by a factor around two, but speeds up calculations by several orders of magnitude for extreme $B$ cases. Nevertheless, we indicate that $\eta_{\rm B}\rightarrow 1$ is too high, as in the region of interest the jet power is likely largely dominated by kinetic energy, and is considered here for illustrative purposes only.

To find the energy evolution up to a given segment, one computes the initial energy $E_0$ that electrons of energy $E$ had when they were injected at the jet base. This is done iteratively by making small time steps back in time (and in space) to sample the energy evolution of electrons backwards from a given segment:
        \begin{equation}\label{E}
        E_{\rm i-1} = E_{\rm i} - \dot{E}(E_{\rm i}) {\rm d}t \ .
        \end{equation}
Here, $E_{\rm i-1}$ is the energy that an electron had at a time $t - {\rm d}t$, and $E_{\rm i}$ is the electron energy at time $t$. We note that $E_{\rm i-1} > E_{\rm i}$ due to the negative sign of $\dot{E}$. The linear approximation used in Eq.~\eqref{E} is only valid as long as $E_{\rm i-1}-E_{\rm i}\ll E_{\rm i}$.

Once $E_0$ is known, the electron energy distribution at the jet base, $N_0(E_0)$, of a segment k with length $dl_{\rm k}$, is computed as
        \begin{equation}\label{NE0}
        N_0(E_0) = Q(E_0) \ t_{\rm k}^{\rm adv} \ ,
        \end{equation}
with $t_{\rm k}^{\rm adv} = dl_{\rm k} / v_{\rm j}$ (if small enough; otherwise $t_{\rm k}^{\rm adv}\rightarrow t_{\rm loss}$). The electron energy distribution at a segment k is computed from the distribution when at the segment k-1:
        \begin{equation}\label{NEk}
        N_k(E_k) = N_{{\rm k}-1}(E_{{\rm k}-1}) \frac{\dot{E}_{\rm k}(E_{{\rm k}-1})}
                   {\dot{E}_{\rm k}(E_{\rm k})} \ ,
        \end{equation}
where $E_{\rm k}$ and $E_{{\rm k}-1}$ are energies related by Eq.~\eqref{E} in segments k and k-1, respectively\footnote{We note that Eq.~(\ref{NEk}) should be corrected for segment length and velocity if they are not constant.}.
By applying this procedure down to the jet base, one gets the general expression for the electron energy distribution at each jet segment:
        \begin{equation}\label{NE}
        N_{\rm k}(E_{\rm k}) = N_0(E_0) \prod_{\rm i=k}^1 \frac{\dot{E}_{\rm i}(E_{{\rm i}-1})}
                               {\dot{E}_{\rm i}(E_{\rm i})} \ ,
        \end{equation}
keeping in mind that $E_{\rm k}\gtrsim m_e c^2$ and $E_0<E_{\rm max}$.

\subsection{Emission of radiation}

The power per photon energy unit radiated by an electron of energy $E$ via synchrotron emission is computed following \citep{pacholczyk70}
        \begin{equation}\label{Psyn}
        P_\varepsilon(E) = \frac{\sqrt{3} e^3 B \sin \theta}{h m_{\rm e} c^2} F(x) \ ,
        \end{equation}
where $\theta$ is the pitch angle between the electron velocity and $B$ vectors, and $x = \varepsilon/\varepsilon_{\rm c}$, with $\varepsilon_{\rm c}$ being the characteristic synchrotron photon energy defined as
        \begin{equation}\label{Ec}
        \varepsilon_{\rm c} = \frac{3}{4\pi} \frac{e h B \sin \theta}{m_{\rm e}^3 c^5} E^2 \ .
        \end{equation}
The function $F(x)$ includes an integral of the $K_{5/3}(\zeta)$ Bessel function, but in this work we adopt an approximation valid within a few percent error in the range $0.1 < x < 10$ \citep[e.g.,][]{melrose80,aharonian04}:
        \begin{equation}\label{Fx}
        F(x) = x \int_x^\infty K_{5/3}(\zeta) \mathrm{d} \zeta \approx 1.85 x^{1/3} e^{-x} \ .
        \end{equation}
The geometry of $\boldsymbol{B}$ is not well known, so an isotropic distribution in the fluid frame is assumed, taking for simplicity $B \sin \theta \approx \sqrt{<B^2>} = B \sqrt{2/3}$. 
The synchrotron spectral energy distribution (SED) for an isotropic population of electrons with distribution $N(E)$ is:
        \begin{equation}\label{Lsyn}
        \varepsilon \ L_\varepsilon^{\rm sync} = \varepsilon \int_0^\infty
        P_\varepsilon(E) N(E) {\rm d}E \ .
        \end{equation}

The IC emission is computed considering that relativistic electrons interact with a beam of photons with energy $\varepsilon_0$. The IC kernel for an angle $\theta$ with respect to the initial beam direction is \citep{aharonian81}:
        \begin{equation}\label{dnIC}
        \frac{{\rm d}^2 N(\theta,\varepsilon)}{{\rm d}\varepsilon \ {\rm d}\Omega} =
    	\frac{3 \sigma_T m_e^2 c^4}{16 \pi \varepsilon_0 E^2} \left[ 1 + \frac{z^2}{2(1-z)} -
    	\frac{2z}{b_\theta(1-z)} + \frac{2z^2}{b_\theta^2(1-z)^2} \right] \ ,
        \end{equation}
where $b_\theta = 2(1-\cos \theta) \varepsilon_0 E/m_{\rm e}^2 c^4$, $z = \varepsilon/E$, and $\varepsilon$ can vary in the limits $\varepsilon_0 < \varepsilon < E b_\theta/(1+b_\theta)$. The (apparent) IC SED at a given direction is obtained by convolving Eq.~\eqref{dnIC} with the electron energy distribution $N(E)$ and the energy distribution density of stellar photons $n(\varepsilon_0)$:
        \begin{equation}\label{LIC}
        \varepsilon \ L_\varepsilon^{\rm IC} = 4 \pi c \varepsilon^2 \int_0^\infty {\rm d}E
    	\int_0^\infty {\rm d}\varepsilon_0 \frac{{\rm d}^2 N(\theta,\varepsilon)}
    	{{\rm d}\varepsilon \ {\rm d}\Omega} N(E) n(\varepsilon_0) \ .
        \end{equation}
Target photons other than those of stellar origin, like those coming from an accretion disk, are unimportant for the jet regions studied in this work. Synchrotron self-Compton is also negligible even when one assumes that most of the nonthermal energy is released as synchrotron radiation.

\subsection{Absorption of radiation}

The main absorption mechanism considered in this work is electron-positron pair production from gamma rays interacting with stellar photons. The cross section for this process is \citep{gould67}
        \begin{equation}\label{gg}
        \sigma_{\gamma\gamma} = \frac{3}{16} \sigma_{\rm T}(1-\beta^2)
    	\left[ (3-\beta^4) \log \left( \frac{1+\beta}{1-\beta} \right) - 2\beta(2-\beta^2) \right] \ ,
        \end{equation}
where
        \begin{equation}\label{beta}
        \beta = \sqrt{1-
        \frac{2 m_{\rm e}^2 c^4}{\varepsilon \varepsilon_0(1-\cos\theta_{\gamma\gamma})}} \ ,
        \end{equation}
$\varepsilon$ and $\varepsilon_0$ are the energies of the high-energy and stellar photons, respectively, and $\theta_{\gamma\gamma}$ is the angle between their propagation directions. 

The optical depth of gamma-ray absorption is \citep{gould67}
        \begin{equation}\label{tau}
        \tau_{\gamma\gamma}(\varepsilon) = \int_0^l {\rm d}l \
    	[1-\cos\theta_{\gamma\gamma}(l)] \int_{\varepsilon_{0,{\rm min}}}^\infty {\rm d}\varepsilon_0 \
    	n(\varepsilon_0,l) \sigma_{\gamma\gamma}(\varepsilon_0,\varepsilon,\theta_{\gamma\gamma})\,,
        \end{equation}
where $l$ is the distance to the observer covered by the gamma ray, and $\varepsilon_{0,{\rm min}} = 2 m_{\rm e}^2 c^4/\varepsilon(1-\cos\theta_{\gamma\gamma})$ is the minimum energy of the gamma rays absorbed by the target stellar photons.

For the cases studied here, synchrotron self-absorption is estimated in radio by computing the optically thick-to-optically thin transition frequency \citep[e.g.,][]{bosch09b}, and is found to be negligible in general; it has therefore not been included in the calculations. Free-free radio absorption in the stellar wind is also negligible at the scales of the emitting region. Nonthermal ultraviolet (UV) photons are also strongly absorbed on their way to the observer, but we have not included this effect as UV photons from the star would be largely dominant in any case. In a very compact system with a very powerful stellar wind, such as, for example, Cygnus~X-3, radio absorption could still be an issue at a distance $D_\star$ from the binary, although we defer specific source studies for future work.

\section{Results}\label{results}

The results are obtained using the parameter values given in Table~\ref{tab:parameters}, together with the sets of values presented in Table~\ref{tab:change} for $i$, the orbital phase $\alpha$, and $\eta_B$, whose variations have strong effects on the nonthermal emission. Considering the free parameters, the electron energy distribution is affected only by $\eta_B$, whereas SEDs, light curves, and the emitter geometry are also affected by $\alpha$ and $i$. Most SEDs are calculated for $\alpha=0.25$ as a typical example, which corresponds to one of the orbit nodes. Varying $\alpha$, we also compute some SEDs and IC light curves to illustrate orbital variations of this component. A comparison of SEDs and light curves is done between the extended and the point-like (one-zone) emitter, to check which features arise when an extended (helical) jet is considered. Finally, maps at 5~GHz are computed to illustrate the expected radio morphologies from a helical jet.

\begin{table}[h]
 \begin{center}
  \begin{threeparttable}
	\caption{Values adopted for different parameters.}
	\begin{tabular}{c c}
    \hline \hline
	Parameter	&	Values									\\
    \hline
    $\eta_B$	&	$10^{-4}$, $10^{-2}$, $1$				\\
    $i$			&	30$^\circ$, 60$^\circ$					\\
    $\alpha$	&	$0-1$									\\
    \hline
    \end{tabular}
    \label{tab:change}
  \end{threeparttable}
 \end{center}
\end{table}

\subsection{Nonthermal electrons}\label{ntel}

\begin{figure}[h]
        \centering
        \includegraphics[width=\linewidth]{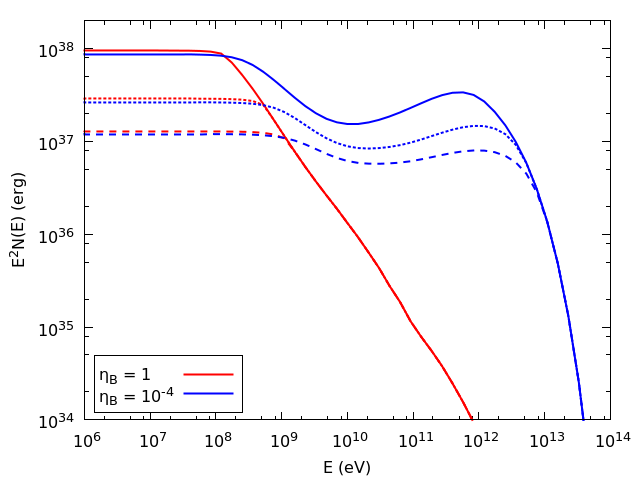}
        \caption{Electron energy distribution of the jet $\times \ E^2$ up to $D_\star$ (dashed lines), $3D_\star$ (dotted lines) and $25D_\star$ (the whole jet; solid lines), for $\eta_B=10^{-4}$ and 1.}
        \label{fig:NE}
\end{figure}

The electron energy distribution for the jet\footnote{The jet and counter-jet electron energy distributions are equal in our model.}, $\times \ E^2$ to emphasize the energy content at different energy scales, is shown in Fig.~\ref{fig:NE} for $\eta_B=10^{-4}$ and 1, and up to three different jet lengths starting from $\boldsymbol{r_0}$: $D_\star$, $3D_\star$ and $25D_\star$ (whole jet). As seen in the figure, a significant fraction of electrons below $\sim 1$~GeV survive beyond $3D_\star$, and for the lowest $B$, TeV electrons can also reach a larger distance from the binary \citep[see also, e.g.,][]{khangulyan08}. As expected, the energy distribution is steep for electron energies dominated by synchrotron and Thomson IC losses. When IC dominates, the SED becomes harder at $\sim 10$~GeV because of the Thomson-to-Klein-Nishina transition of the electron distribution, getting steeper again only at the highest energies due to synchrotron losses (and the effect of $E_{\rm cut}$).

\subsection{Spectral energy distribution}

\begin{figure*}
        \centering
        \includegraphics[width=0.48\linewidth]{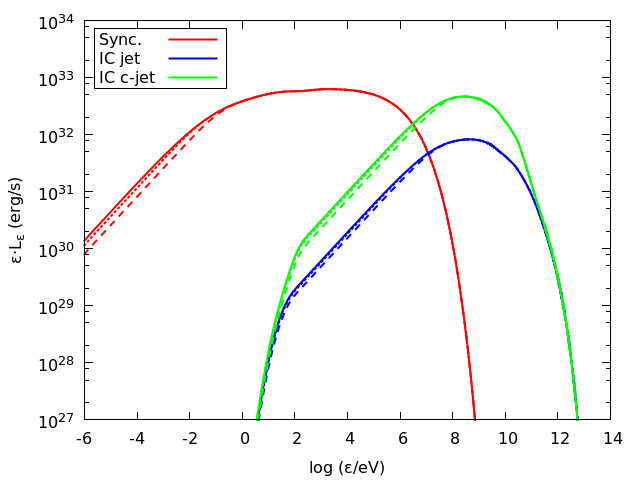}
    	\includegraphics[width=0.48\linewidth]{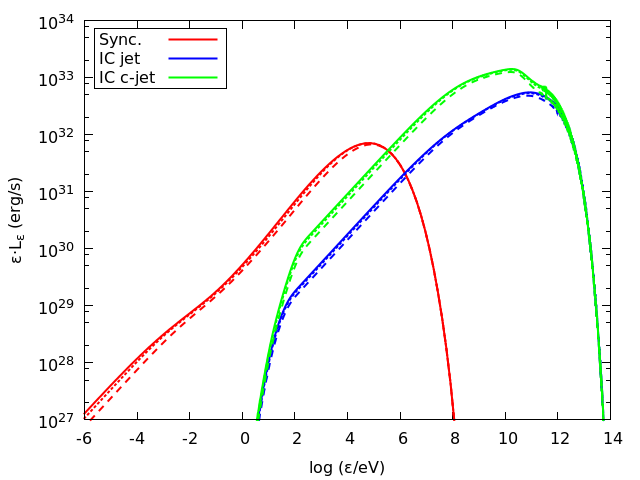}
    	\caption{Synchrotron and IC SEDs of the jet (blue) and the counter-jet (green), for $i$ = $60^\circ$ and $\alpha=0.25$, up to $D_\star$ (dashed lines), $3D_\star$ (dotted lines), and $25D_\star$ (the whole jet; solid lines), for $\eta_B = 10^{-4}$ (right panel) and 1 (left panel).}
        \label{fig:SEDpt}
\end{figure*}

Figure~\ref{fig:SEDpt} shows the SEDs of the synchrotron and the IC emission for the same cases studied in Fig.~\ref{fig:NE}, showing now within each panel the jet and the counter-jet contributions separately. An inclination of $i=60^\circ$ is adopted, and $\alpha=0.25$. As expected from the severe drop in radiation efficiencies, the contribution to the emission of the regions beyond $3D_\star$ is minor for all the cases studied, although radio emission is still non-negligible beyond that distance, as seen in Sect. \ref{radio}. It is worth noting that most of the IC radiation comes from the counter-jet because of the larger IC $\theta$-values. Doppler boosting is negligible for the adopted $v_{\rm j}$-value and is not included.

\begin{figure}
        \centering
        \includegraphics[width=\linewidth]{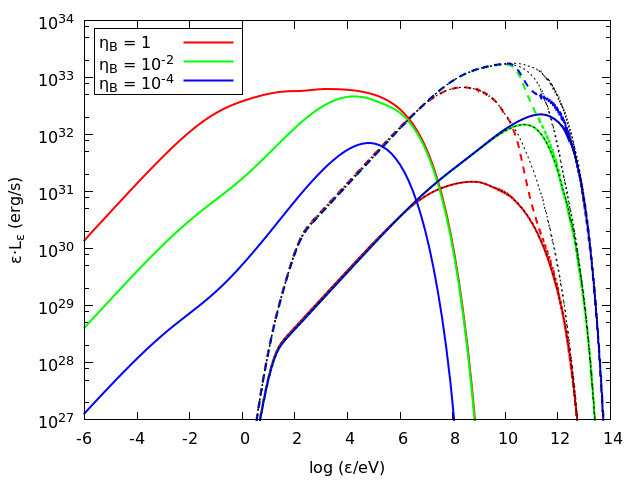}
        \caption{Synchrotron and IC SEDs of the jet (solid lines) and the counter-jet (dashed lines), for $i=30^\circ$ and $\alpha=0.25$, and $\eta_{\rm B}=10^{-4}$ (blue lines), $10^{-2}$ (green lines), and 1 (red lines). The unabsorbed IC emission is also shown (black dotted lines).}
        \label{fig:SEDB}
\end{figure}

\begin{figure}
        \centering
        \includegraphics[width=\linewidth]{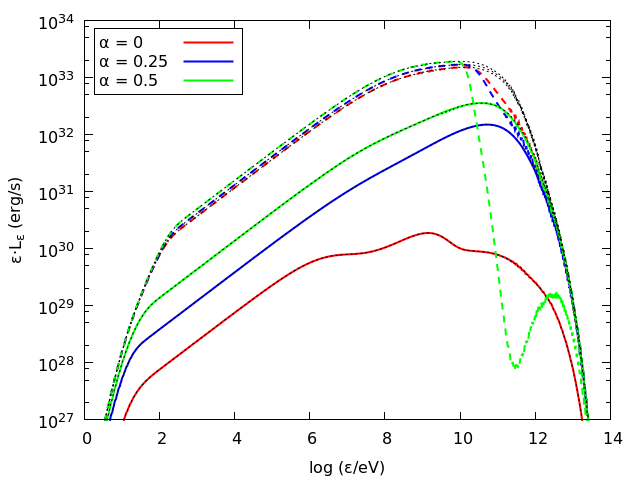}
        \caption{IC SEDs of the jet (solid lines) and the counter-jet (dashed lines), for $i=30^\circ$, $\eta_{\rm B}=10^{-2}$, $\alpha=0$ (red lines), 0.25 (blue lines) and 0.5 (green lines). The unabsorbed IC emission is also shown (black dotted lines).}
        \label{fig:SEDOrb}
\end{figure}

The effect of the magnetic field on the synchrotron and IC SEDs is shown in Fig.~\ref{fig:SEDB} for $i=30^\circ$ and $\alpha=0.25$, and for the jet and the counter-jet separately. As expected, the synchrotron to IC luminosity ratio depends strongly on $\eta_{\rm B}$. Figure~\ref{fig:SEDB} also shows the effect of gamma-ray absorption (stronger for $i=30^\circ$ than for $60^\circ$), which is quite modest given the relatively large distances from the star and the $\alpha$-value considered in this plot.

Figure \ref{fig:SEDOrb} shows the effect of the orbital phase on the IC spectrum. With $i=30^\circ$ and $\eta_B=10^{-2}$, the SEDs for $\alpha=0, 0.25$ and $0.5$ are plotted. For $\alpha=0.5$, which corresponds to the CO superior conjunction (i.e., the CO is behind the star), there is a drop of up to five orders of magnitude in the counter-jet emission above 10~GeV due to gamma-ray absorption, which makes the jet dominate the radiation output in this range, effectively smoothing the overall impact of gamma-ray absorption for this orbital phase.

\subsection{Light curve}

\begin{figure*}
        \centering
        \includegraphics[width=0.49\linewidth]{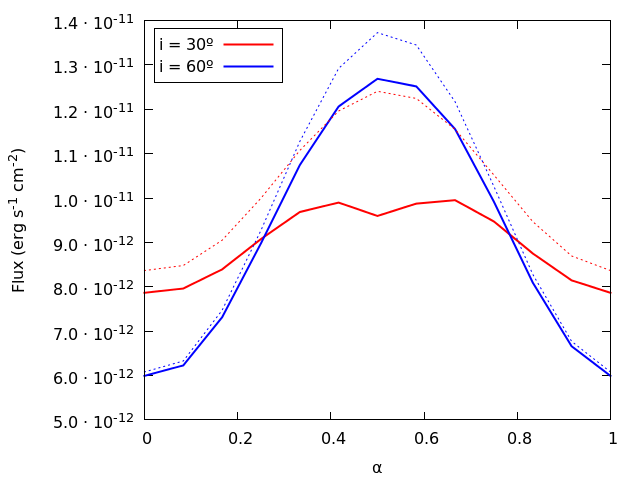}
    	\includegraphics[width=0.49\linewidth]{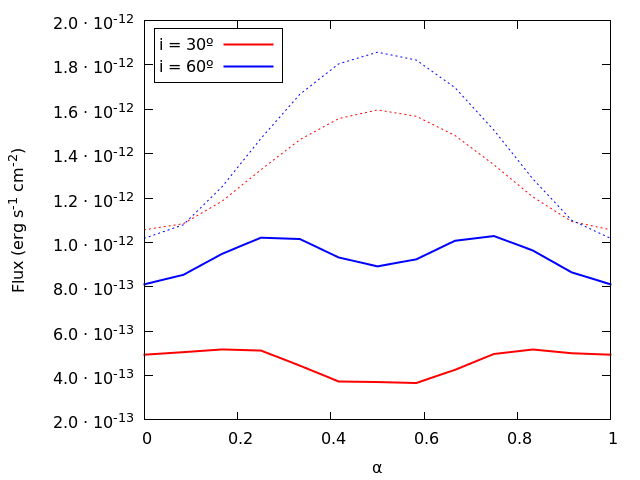}
        \caption{Light curves of IC emission (jet + counter-jet) at 0.1--100~GeV (left panel) and $>100$~GeV (right panel), for $i=30^\circ$ (red lines) and $60^\circ$ (blue lines), and $\eta_B=10^{-2}$. The unabsorbed light curves are also shown (dotted lines). The CO is in inferior (superior) conjunction with the star when the orbital phase is 0 (0.5).}
        \label{fig:Flux}
\end{figure*}

Figure~\ref{fig:Flux} shows the IC light curves for $\eta_B = 10^{-2}$. The peak around $\alpha=0.5$ and $0.1-100$~GeV for $i=60^\circ$ is explained by the larger IC $\theta$-angles, and the small dip for $i=30^\circ$ manifests some level of gamma-ray absorption, which starts at $\sim 10$~GeV in a relatively hard IC SED (see Fig.~\ref{fig:SEDOrb}). For energies $>100$~GeV, absorption has a very strong impact on the fluxes by a factor of several, although the jet component smoothens somewhat the light curve around $\alpha=0.5$ (superior conjunction of the CO) because of reduced gamma-ray absorption due to a smaller $\theta_{\gamma\gamma}$-value (see Fig.~\ref{fig:SEDOrb}). On the other hand, the counter-jet still leads to significant IC fluxes around $\alpha=0$ (inferior conjunction; see Fig.~\ref{fig:SEDOrb}), unlike an emitter located very close to the CO, which would lead to stronger differences in flux.

\subsection{Extended versus one-zone emitter}

\begin{figure}
        \centering
        \includegraphics[width=\linewidth]{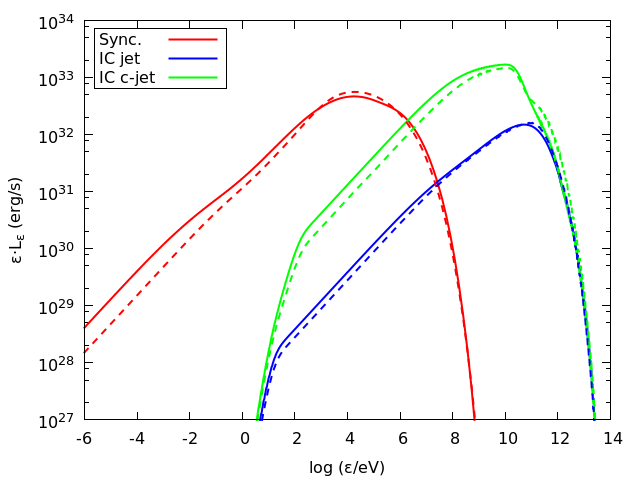}
        \caption{Comparison of the (absorbed) SEDs of an extended (helical) jet (solid lines) and a one-zone emitter (dashed lines) for the jet and the counter-jet for $i$ = 30$^\circ$, $\alpha=0.25$, and $\eta_B = 10^{-2}$.}
        \label{fig:SED1Z}
\end{figure}

Figure~\ref{fig:SED1Z} shows a comparison between the SEDs of an extended (helical) jet and a one-zone emitter, for the jet and the counter-jet, for $i=30^\circ$, $\alpha=0.25$, and $\eta_B = 10^{-2}$. The one-zone emitter is assumed to be just one segment located at $\boldsymbol{r_0}$ and of length $D_\star/3$, which sets the electron escape time. Because of fast radiation losses, the one-zone approximation works well in the high-energy part of the synchrotron and IC SEDs, as expected. Qualitatively, the extended and the one-zone emitters look quite similar, although differences of up to a factor of 2 are seen, mainly but not only in the jet synchrotron and IC components at low energies, as low-energy electrons escaped from the one-zone region are still radiatively relevant. We note that inaccurate source knowledge and model simplifications imply systematic uncertainties likely larger than a factor of 2, which means that the SED alone cannot help to discriminate a helical jet model from that of a one-zone emitter.

\begin{figure*}
        \centering
        \includegraphics[width=0.48\linewidth]{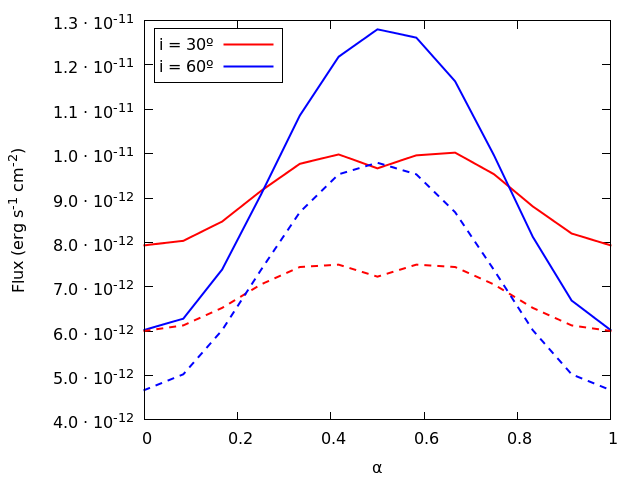}
    	\includegraphics[width=0.48\linewidth]{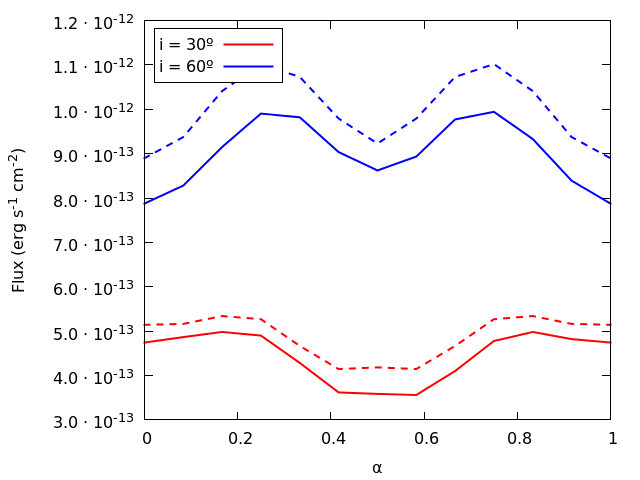}
        \caption{As in Fig.~\ref{fig:Flux} but comparing the extended jet (solid lines) and the one-zone case (dashed lines), and for the absorbed gamma-ray component only.}
        \label{fig:Flux1Z}
\end{figure*}

Figure~\ref{fig:Flux1Z} compares the light curves of an extended and a one-zone emitter in the cases presented in Fig.~\ref{fig:Flux}. The fluxes are larger (lower) for the extended emitter for $0.1-100$~GeV ($>100$~GeV), and are symmetric around $\alpha=0.5$. The different behavior between the two energy bands is caused by the fact that electrons emitting $>100$~GeV photons via IC reach farther from the star than those emitting $0.1-100$~GeV photons (see Sect.~\ref{ntel}), both for the counter-jet and the jet, decreasing the IC flux as they meet a more diluted target photon field. We note that for higher jet powers, the helical jet would start further away (see Eqs.~\ref{chi} and \ref{xturn}), in which case the light curve would become asymmetric for the extended jet.

\subsection{Radio emission}\label{radio}

\begin{figure*}
        \centering
        \includegraphics[width=\linewidth]{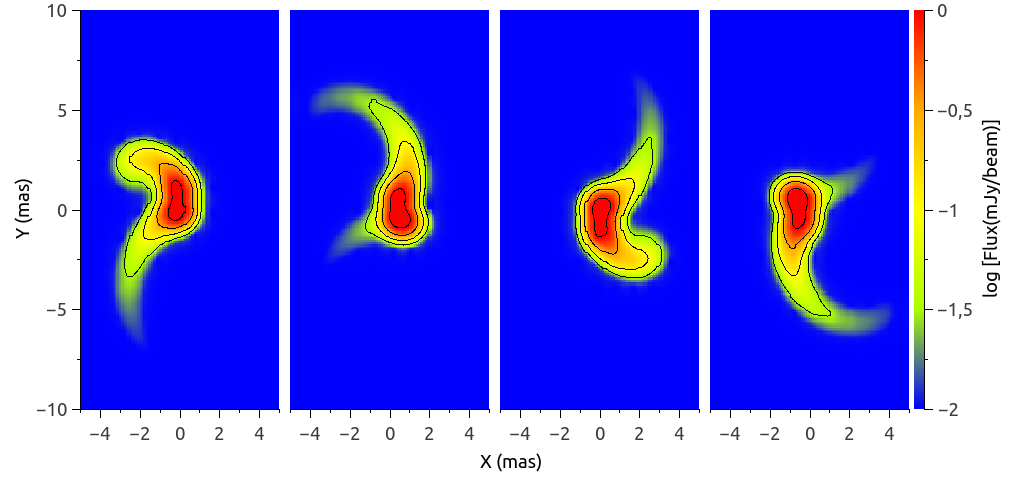}
        \caption{Sky maps at 5 GHz for $i$ = 30$^\circ$, $\eta_B = 10^{-2}$, and orbital phases of 0, 0.25, 0.5 and 0.75 from left to right. Contour levels correspond to fluxes of 0.032, 0.1, 0.32 and 1 mJy/beam. The beam size is 1~mas.}
        \label{fig:Sky30}
\end{figure*}

\begin{figure*}
        \centering
        \includegraphics[width=\linewidth]{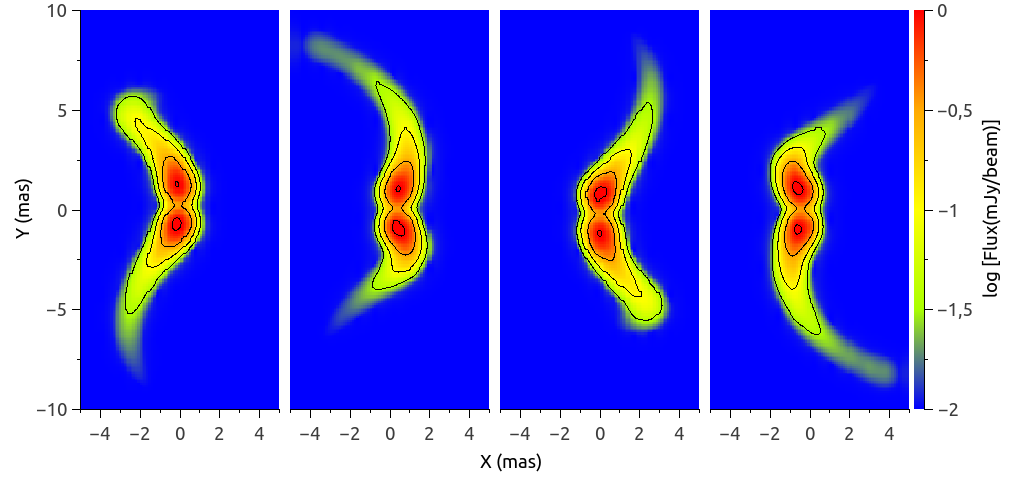}
        \caption{As in Fig. \ref{fig:Sky30} but for $i$ = 60$^\circ$.}
        \label{fig:Sky60}
\end{figure*}

\begin{figure}
        \centering
        \includegraphics[width=\linewidth]{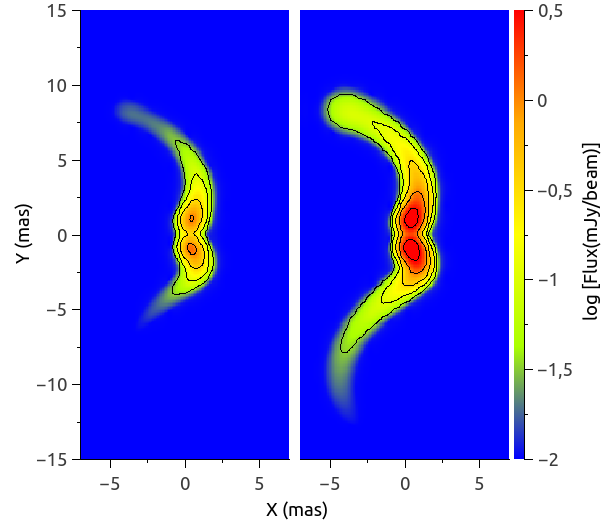}
        \caption{Sky maps at 5 GHz for $i=60^\circ$, $\alpha=0.25$, $\eta_{\rm NT} = 10^{-2}$ (left) and $5\times10^{-2}$ (right). Contour levels are (mJy/beam): 0.032, 0.1, 0.32, 1, 3.2. The beam size is 1~mas.}
        \label{fig:SkyNT}
\end{figure}

Radio sky maps at 5 GHz for $\eta_B = 10^{-2}$ and different orbital phases are shown in Figs. \ref{fig:Sky30} and \ref{fig:Sky60} for $i$ = 30$^\circ$ and $i$ = 60$^\circ$, respectively. These maps are obtained following three steps: first, the helical jet structure is projected into the plane of the sky. Subsequently, the emission from each segment is convolved with a Gaussian with standard deviation $\sigma = r_{\rm j}/2$ to approximately distribute the emission over the actual sky-projected area of each segment. Finally, this emission is convolved again with a Gaussian with FWHM = 1~mas in order to mimic the response of a radio, very long baseline interferometer \citep[see, e.g.,][for a VLBA description]{walker95}. With the sensitivity of current instrumentation, typically of a few tens of $\upmu$Jy/beam, the extended emission could be resolved, but only from the parts of the jet that are relatively close to its base. The bending of the jet appears more dramatic the smaller the inclination due to projection effects. The total received flux for each map is 7.4~mJy. As the flux increases proportionally to $\eta_{\rm NT}$, one could detect the jet up to larger distances for higher $\eta_{\rm NT}$. To illustrate this, Fig. \ref{fig:SkyNT} shows a comparison between two sky maps with different nonthermal energy fractions, $\eta_{\rm NT}=10^{-2}$ and $\eta_{\rm NT}=5\times10^{-2}$. We note that the radio fluxes also depend on $\eta_B$ as $\propto \eta_B^{3/4}$.

\section{Summary and discussion}\label{discussion}

As argued by \cite{bosch16}, a nonballistic helical jet region is the likely outcome of a HMMQ jet interacting with the stellar wind under the effect of orbital motion. In the present work, we have shown that such a region is predicted to produce significant fluxes from radio to gamma rays if $L_{\rm NT}\sim \eta_{\rm NT}L_{\rm j}\gtrsim 10^{34}\eta_{\rm NT,-2}$~erg~s$^{-1}$, for a source at a few kiloparsecs. These significant fluxes are explained by the high synchrotron and IC efficiencies at the helical jet onset location, $\boldsymbol{r_0}$, for typical HMMQ $L_\star$-values. Since realistic $\eta_B$-values are expected to be well below equipartition with the radiation field, the IC component is likely to dominate the nonthermal emission, peaking around 10~GeV. Specific gamma-ray light curve features are also predicted: a non-negligible impact of gamma-ray absorption, combined with angular effects for this process and IC. Peculiar changing radio morphologies, which can trace the jet helical structure, are expected as well.

We note that hydrodynamical instabilities and wind-jet mixing make our results mostly valid for the inner helical jet region, say up to a few $D_\star$. Beyond that point, the jet helical geometry is likely to become blurred, eventually turning into a bipolar, relatively wide, and collimated supersonic outflow; a mixture of jet and wind material. Although the flow may still be mildly relativistic for very energetic ejections, in general the resulting "jet" is expected to be much faster than the stellar wind, but nonrelativistic. 

Our results are based on a number of strong simplifications, in particular regarding the stability of the helical jet. Therefore, detailed numerical simulations of the jet-wind interaction on middle to large scales are necessary for more accurate predictions. Nevertheless, despite the simplifications adopted, the main features of the scenario are expected to be rather robust, at least at a semi-quantitative level. These are: non-negligible fluxes; specific light curves characterized by the value of $\boldsymbol{r_0}$, the interplay of jet and counter-jet emission, and angular IC and gamma-ray absorption effects; and curved jet radio morphologies with characteristic orbital evolution. This is so because the first- or even zeroth-order level of the dynamical, radiative, and geometrical effects involved are taken into account in our calculations.

Two important system parameters, eccentricity and $d_{\rm O}$, have been assumed constant in our calculations. Despite being difficult to ascertain the impact of eccentricity in a high-mass microquasar without numerical calculations, simulations of pulsar-star wind colliding in eccentric binaries \citep{barkov16,bosch17} suggest that the effect of eccentricity should be important only for rather eccentric binaries. In that case, the nature of the interaction structure may not be helical at all, being not only strongly asymmetric, and inclined towards the apastron side of the orbit, but also likely much more sensitive to disruption. The value of $d_{\rm O}$, and thus $P$, may also change. This would leave the helical jet geometry unaffected: both the helical vertical step and $x_{\rm turn}$ are $\propto P\propto d_{\rm O}^{3/2}$. On the other hand, the luminosity would change as $\propto P^{-1}\propto d_{\rm O}^{-3/2}$ except for emission produced under the fast radiative cooling regime of electrons (in our setup the highest-energy end of the synchrotron and IC emission), which is independent of $d_{\rm O}$. The helical jet radiation in a very wide system would be hardly detectable unless the flow is slow, which allows electrons to radiate more energy. On the other hand, in a very compact system, radio emission would be hard to resolve for a slow helical jet flow, as the source would be too small.

The emission on the scales of the binary, produced for instance by jet internal shocks or by the first wind-induced recollimation shock in the jet (stage 2), could be important. In which case: (i) assuming that the same nonthermal luminosity is injected in both regions, the radio emission from the binary scales would be strongly absorbed via synchrotron self-absorption and wind free-free absorption; and (ii) the X- and gamma-ray emission would be at a similar level if produced in the fast radiation cooling regime, and at a higher level otherwise because of the higher $B$-value and IC-target density. Gamma-ray absorption could however strongly attenuate the gamma-ray luminosity $>100$~GeV for most of the orbit. Therefore, the helical jet could contribute significantly to the overall nonthermal radiation of HMMQ, and particularly in radio and $>100$~GeV, even if electrons are also accelerated closer to the jet base. This was the main motivation for this work: to perform a first exploration of specific radiation features of the helical jet region, so that it could be disentangled from the other emitting sites. For the HMMQ Cygnus~X-1 and Cygnus~X-3, their uncertain wind and jet parameters make it difficult to make concrete predictions; for example, a non-ballistic jet region may not form at all \citep[see][and references therein]{yoon16,bosch16}, and an individual detailed source analysis is out of the scope of this work. In spite of this, our findings are not in contradiction with the radio morphology and the gamma-ray light curves of these sources \citep[e.g.,][]{stirling01,mioduszewski01,miller04,zanin16,zdziarski18}. Future detailed modeling together with deep, high-resolution radio observations and the high-energy light curves could provide hints of helical jet emission.

The wind-jet interaction in HMMQ resembles, to a significant extent, the wind-wind interaction in high-mass binaries hosting young pulsars. The latter sources are probably extended mostly on the orbital plane and along the orbit semi-major axis, whereas the bipolar helical-jet structure is focused mostly in a direction perpendicular to the orbit plane \citep{bosch15,bosch16,bosch17}. Nevertheless, the interactions of the stellar wind with a relativistic pulsar wind or jet, on binary, middle, and large scales, share many qualitative properties, and this may mask a fundamentally different engine (accretion vs. a pulsar wind) when observing sources of yet unknown CO. These similarities are: extended radio structures with orbital evolution found at mas scales; strong radio absorption on the binary scales; similar hydrodynamics, and therefore impact of instabilities, adiabatic losses, and Doppler boosting effects\footnote{The much higher energy per particle of shocked pulsar winds can lead to stronger Doppler boosting effects than in a HMMQ jet, although mass-loading smoothens this difference.}, for the shocked flows outside the binary region; and gamma-ray light curves strongly affected by IC and gamma-ray absorption angular effects, in a likely extended/multi-zone emitter. Radio observations, high-quality multiwavelength data, numerical simulations, and realistic radiation calculations \citep[e.g.,][]{hess06,perucho08,magic09,fermi09,perucho10,moldon11a,moldon11b,moldon12,perucho12,bosch15,dubus15,zanin16,yoon15,yoon16,delacita17,bosch17,wu18,barkov18} are therefore needed to characterize radiation scenarios in high-mass binaries with unknown CO, such that empirical observables can help to disentangle the CO nature.

\begin{acknowledgements}
We would like to thank the referee for a constructive and useful report that helped to improve the manuscript. We acknowledge support by the Spanish Ministerio de Econom\'{i}a y Competitividad (MINECO/FEDER, UE) under grants AYA2013-47447-C3-1-P and AYA2016-76012-C3-1-P, with partial support by the European Regional Development Fund (ERDF/FEDER), MDM-2014-0369 of ICCUB (Unidad de Excelencia `Mar\'{i}a de Maeztu'), and the Catalan DEC grant 2017 SGR 643. EM acknowledges support from MINECO through grant BES-2016-076342.
\end{acknowledgements}

\FloatBarrier
\bibliographystyle{aa}
\bibliography{references}

\end{document}